# Distributed Heterogeneous Relational Data Warehouse In A Grid Environment

Saima Iqbal, Julian J. Bunn, Harvey B. Newman
*California Institute of Technology, Pasadena, CA 91125, USA*

This paper examines how a "Distributed Heterogeneous Relational Data Warehouse" can be integrated in a Grid environment that will provide physicists with efficient access to large and small object collections drawn from databases at multiple sites. This paper investigates the requirements of Grid-enabling such a warehouse, and explores how these requirements may be met by extensions to existing Grid middleware. We present initial results obtained with a working prototype warehouse of this kind using both SQLServer and Oracle9i, where a Grid-enabled web-services interface makes it easier for web-applications to access the distributed contents of the databases securely. Based on the success of the prototype, we proposes a framework for using heterogenous relational data warehouse through the web-service interface and create a single "Virtual Database System" for users. The ability to transparently access data in this way, as shown in prototype, is likely to be a very powerful facility for HENP and other grid users wishing to collate and analyze information distributed over Grid.

## 1. INTRODUCTION

This paper proposes a Web Services architecture based on Grid Services [1]. The concept of this architecture is to make "Distributed Heterogeneous Relational Data Warehouse" (DHRD) databases available across the Grid in the form of Grid Services with the help of Web Services technology.

This architecture implements Grid Database Service Specification [2]. In addition to these specifications, this architecture also provides a JAX-RPC [3] client to access DHRD across the Grid, MonALISA [4] integration to find optimal network resource path for the connection with DHRD and a private UDDI [5] registry server using Xindice for the repository.

In the first test of this architecture prototype two questions addresses: how a client can find the location of a database for a required dataset? How a UDDI registry server can be used for the registry and repository purpose of Grid Services.

## 2. GRID SERVICES

Grid Service is a Web Service that conforms to a set of conventions (interfaces and behaviors) that define how a client interacts with services available across the Grid [6].

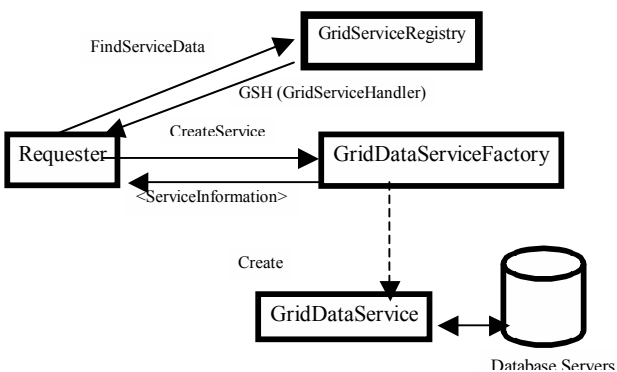

Figure 1: Creating a Grid Data Service

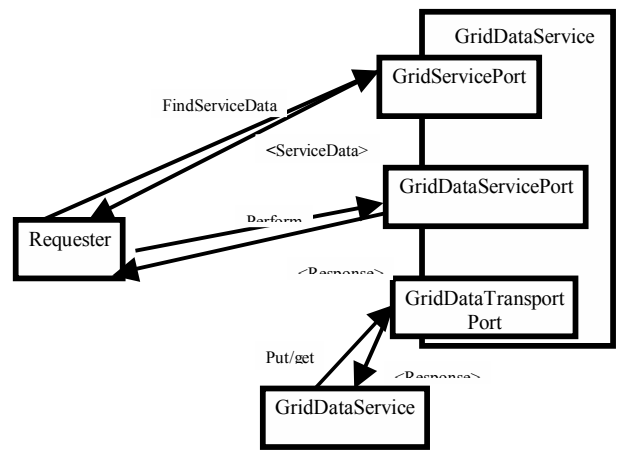

Figure 2: Requester Using Grid Data Service

The Open Grid Service Architecture (OGSA) [7] extends Web Services with consistent interfaces for creating, managing and exchanging information among Grid Services, which are dynamic computational artifacts casts as Web Services. Both the Web and Grid Services communities stand to benefit from provision of consistent, agreed service interfaces to database management systems (DBMS). Such interfaces are required to support the description and use of database systems (DBS) using Web Services standards, taking account of the design conventions and mandatory features of Grid Services.

## 3. WEB SERVICES

*"Web Services are modular software components wrapped inside a specific set of Internet communication protocols and that can be run over the internet".*

At the heart Web Services architecture is the need of program-to-program communications, which is also the requirement of Grid Services. Key roles of the Web Services architecture are: a service provider, a service registry and a service requestor. Together they perform three operations of publish, find and bind on Web Services. These operations are performed with the help of three vital components of Web services: Simple Object Access Protocol (SOAP), Web Services Description Language (WSDL) [8] and Universal Description Discovery and Integration (UDDI) technology.

## 4. INTEGRATION OF DHRD WITH THE GRID

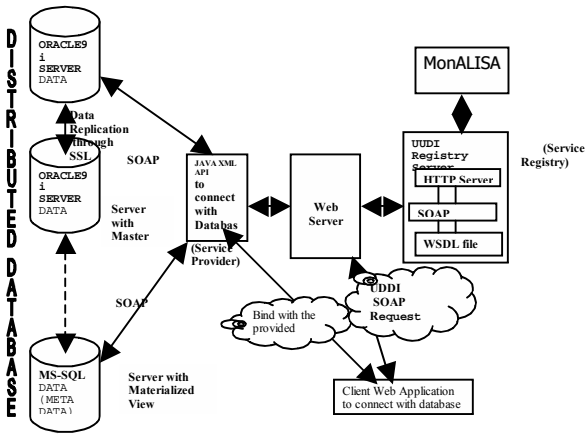

Figure 3: DHRD Web Services Architecture for the Grid

As neither current Grid software, nor existing DBMS are able to fully support the integration of DHRD with the Grid applications, this proposed architecture examines how this can be achieved.

We believe that, it would be a better approach to use Web Services driven design. Each DBS ( here DBS is used rather than DBMS, as the owner of data warehouse can choose which services to make available on the Grid and who is allowed to access them [9] ) would offer a set of services covering the requirements of client application. Individual operations offered by these services would be standardized to increase portability. Where possible reduce the effort required to build applications that interact with multiple DBS. This would be done by adding code to map the operation interface offered by a service to the vendor specific interface beneath. However, it is impossible to standardize all services: for example different database paradigms support different types of query languages and these can not all be reconciled into a standard query language (even within relational databases there are variations). One advantage of the web service driven approach is however that each DBS made available on the Grid can provide a metadata service in form of WSDL that gives information on the range of services and operations that it supports. This would give application developers the information required to exploit whatever facilities are available.

The architecture shown in Figure 3 shows the Web Services based approach in which DBS, across the Grid,

register themselves in UDDI registry server as a web service. The client, a service requester, sends a request for a required dataset to the web server. On the basis of the information provided by the request of a client, server-classes query the UDDI registry server. This query provides the endpoint of the required DBS service and will also use the network information provided by MonALISA, to find the optimal network connection to connect with the required database. Finally, the required service endpoint is available to the client request which bind the client application with the required database, to perform transactions on DBS.

## 5. PROTOTYPE TEST

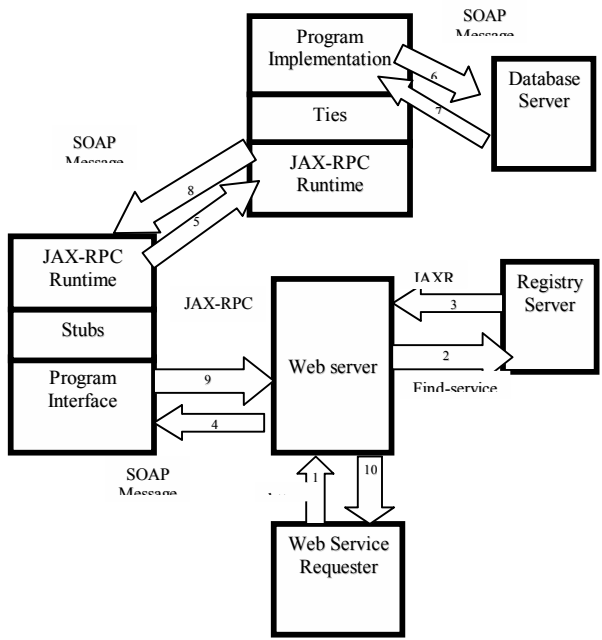

Figure 4: Web Services prototype at runtime

### 5.1. Technologies Employed

In this prototype JAX-RPC [3], stands for Java API for XML-based RPC, is used to create Web services and client that use Remote Procedure Calls (RPC) and XML. JAXR [10], Java API for XML registries, used to register and discover the services. Java Web Services Developer Pack (JWDP [11]) registry server 1.0_02, implements version 2.00 of UDDI and use Xindice database for the repository of the registry data. Xrpcc [12] tool to generate WSDL. Apache web server [13] and Tomcat servlet engine.

### 5.2. Working

Figure 5 shows a simplified view of the DBS service after it has been deployed. After receiving the client's http request web server uses JAXR to send a query to the UDDI registry server. Then search for the required service (read WSDL) that supports JAX-RPC. After having the URL of **endpoint,** implementation class of the client program on web server invokes a method on the **stub**. The stub invokes the routines in the **JAX-RPC runtime system**. The runtime system converts the remote method call into a **SOPA** message as an **HTTP** request. When the server receives the **HTTP** request, the **JAX-RPC runtime system** extracts the **SOAP** message from the request and then translates it into the method call. The **JAX-RPC runtime system** invokes the method on the **tie** object. The **tie** object invokes the method on the implementation of the **DBS service**. The run time system on the server converts the method's response into SOAP message and then transmits the message back to the client as an **HTTP** response. On the client side the **JAX-RPC runtime system extracts the SOAP message from the HTTP response and then translates** it into a method response for the client program.

### 5.3. Screen Shots of Prototype

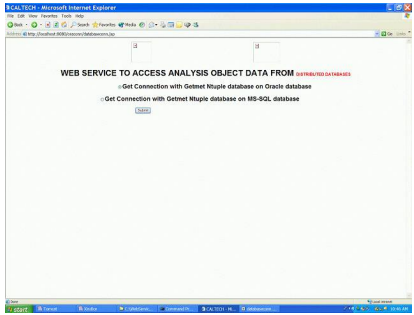

Figure 5: JSP page to locate service

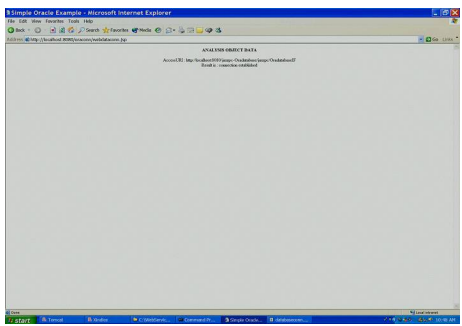

Figure 6: Response of the request: endpoint URL and database connection result

### 6. FUTURE WORK

Future development of this prototype includes the integration of MonALISA (Grid Monitoring Tool) and will try to exploit the UDDI features for the solution of Grid Services registry. Develop an API to integrate this Grid Services based Web Services prototype into the Globus toolkit.

### Acknowledgments

This work is supported by CALTECH (California Institute of Technology), USA.